\documentclass[12pt,preprint]{aastex}
\shorttitle{Be Star Spectral Energy Distributions}
\shortauthors{Touhami et al.}
\begin{document}

\received{2009 Aug 31}
\accepted{}

\title{Spectral Energy Distributions of Be and Other Massive Stars} 

\author{Y. Touhami\altaffilmark{1}, N. D. Richardson\altaffilmark{1,2},
D. R. Gies\altaffilmark{1,2,3,4}, G. H. Schaefer\altaffilmark{5}, 
T. S. Boyajian\altaffilmark{1,3}, \\ S. J. Williams\altaffilmark{1,3},
E. D. Grundstrom\altaffilmark{3,4,6},  M. V. McSwain\altaffilmark{3,7},
D. P. Clemens\altaffilmark{8}, and B. Taylor\altaffilmark{8}}

\altaffiltext{1}{Center for High Angular Resolution Astronomy, 
Department of Physics and Astronomy, 
Georgia State University, P. O. Box 4106, Atlanta, GA  30302-4106; 
yamina@chara.gsu.edu, richardson@chara.gsu.edu, gies@chara.gsu.edu,  
tabetha@chara.gsu.edu, swilliams@chara.gsu.edu} 
 
\altaffiltext{2}{Visiting Astronomer, Lowell Observatory, 
supported by a Program for Research and Education with Small Telescopes 
(PREST) grant from the National Science Foundation to Lowell Observatory and
to Boston University.} 
 
\altaffiltext{3}{Visiting Astronomer, Kitt Peak National Observatory, 
National Optical Astronomy Observatory, operated by the Association 
of Universities for Research in Astronomy, Inc., under contract with 
the National Science Foundation.} 
 
\altaffiltext{4}{Visiting Astronomer at the Infrared Telescope Facility, 
which is operated by the University of Hawaii under Cooperative Agreement 
no.\ NCC 5-538 with the National Aeronautics and Space Administration, 
Science Mission Directorate, Planetary Astronomy Program.}
 
\altaffiltext{5}{CHARA Array of Georgia State University, 
Mount Wilson Observatory, Mount Wilson, CA 91023;
schaefer@chara-array.org}

\altaffiltext{6}{Physics and Astronomy Department,  
Vanderbilt University, 6301 Stevenson Center, Nashville, TN 37235;
erika.grundstrom@vanderbilt.edu} 
 
\altaffiltext{7}{Department of Physics, Lehigh University, 
16 Memorial Drive East, Bethlehem, PA 18015; \\ mcswain@lehigh.edu}

\altaffiltext{8}{Institute for Astrophysical Research, 
Boston University, 725 Commonwealth Ave., Boston, MA 02215; clemens@bu.edu, taylor@lowell.edu}

\slugcomment{Submitted to PASP}
\paperid{350326}


\begin{abstract}
We present spectrophotometric data from 0.4 to 4.2 $\mu$m 
for bright, northern sky, Be stars and several other 
types of massive stars.  Our goal is to use these data with 
ongoing, high angular resolution, interferometric observations to model 
the density structure and sky orientation of the gas surrounding
these stars.  We also present a montage of the 
H$\alpha$ and near-infrared emission lines that form 
in Be star disks.  We find that a simplified measurement 
of the IR excess flux appears to be correlated with 
the strength of emission lines from high level transitions
of hydrogen.  This suggests that the near-IR continuum
and upper level line fluxes both form in the inner part of 
the disk, close to the star. 
\end{abstract}



\setcounter{footnote}{8}

\section{Introduction}                              

The observed absolute flux from an astronomical source 
(after correction for telluric and interstellar extinction) is 
directly related to its emitted flux and angular size 
in the sky.  As we enter the era of optical long-baseline 
interferometry, it will become easier to measure the 
angular dimensions of many objects and, consequently,
to explore the relationship between the observed and 
emitted flux distributions.  This effort is especially 
important to determine effective temperatures of stars, 
but it also plays a key role in the interpretation
of circumstellar environments, in particular the disks 
surrounding Be stars and the winds and outflows of massive stars.

Be stars are rapidly rotating B-type stars that manage 
to eject gas into a circumstellar disk (observed in H emission 
lines, an infrared flux excess, and linear polarization; 
\citealt{por03}).  The IR flux excess from the disk results
from bound-free and free-free emission from ionized gas,
and this emission increases with wavelength, 
so that in the near and mid-IR the disk 
flux will dominate over the stellar flux.  Models of 
the IR excess can relate the observations to the disk 
radial density function \citep{wat86,dou94,por99}.
Such models are also required to interpret recent 
near-IR interferometric observations of Be stars 
where the ratio of disk to stellar flux is a 
key parameter \citep{ste01,gie07,mei07,car09}.
However, Be star disks are intrinsically variable on 
timescales of months to years \citep{hub98,por03,mcs08}, so it
is necessary to obtain contemporaneous spectrophotometry
in order to model both the total flux and its 
angular distribution in the sky.   There are many emission 
lines of H and He in the near-IR spectra of Be stars 
\citep{cla00,stl01,len2a,men09}, and they offer 
additional diagnostics of the disk density, temperature, 
and geometry \citep{hon00,len2b,jon09}.   

We have embarked on a number of programs of interferometry 
with the Georgia State University Center for High Angular 
Resolution Astronomy (CHARA) Array, a six-telescope, 
optical/IR interferometer with baselines up to 330 m 
\citep{ten05}.  Here we describe a program of complementary 
optical and near-IR spectrophotometry of our targets 
that we will use in detailed modeling of the source 
angular flux distribution.  
The observations and their calibration 
are described in \S2, and we present figures of the target 
spectral energy distributions and emission line strengths
in \S3 and \S4, respectively.  In \S5 we discuss the relationship 
between the disk continuum and line emission of Be stars, and
we present a summary of the work in \S6. 


\section{Observations and Reductions}               

We obtained optical spectrophotometry in 2006 and 2008 
with NOAO Kitt Peak National Observatory Coud\'{e} Feed telescope. 
These observations cover parts of the blue and red spectrum 
as outlined in Table 1 that lists (1) the UT dates and (2) heliocentric
Julian dates of observation, (3) the wavelength range recorded, 
(4) the spectral resolving power (for a nominal projected slit 
equivalent to three pixels), (5) the number of spectra made, and
(6) a summary of the telescope, spectrographic grating, and 
detector.  The KPNO observations were made with a slit width of 
$4\farcs3 - 9\farcs0$ to record most to all of the starlight,
so the actual spectral resolution depends on the 
stellar point spread function and image motion 
during the exposure.  Each observation was immediately 
preceded or followed by an identical observation of a flux 
calibrator star (discussed further below).  All the observations
were accompanied by dark, flat field, and ThAr comparison
lamp (for wavelength calibration) frames, and the spectra 
were reduced and extracted by standard means in 
IRAF\footnote{IRAF is distributed by the National Optical Astronomical
Observatory, which is operated by the Association of Universities for
Research in Astronomy, Inc. (AURA), under cooperative agreement with the
National Science Foundation.} to produce a spectrum of integrated counts 
per second as a function of heliocentric wavelength. 

\placetable{tab1}      

We also obtained near-IR spectroscopy of most of the targets in 
2006 with the NASA Infrared Telescope Facility and 
SpeX cross-dispersed spectrograph \citep{ray03}
and in 2008 with the Mimir camera/spectrograph and 
Lowell Observatory Perkins Telescope \citep{cle07}.  
Both sets of observations were made with a wide slit 
to accommodate most of the stellar flux ($3\farcs0$ and $10\farcs0$
for SpeX and Mimir, respectively), although we also 
obtained a set of narrow slit ($0\farcs3$), high resolution spectra
with SpeX.  The SpeX data cover the photometric $K$ and $L$ bands
while the Mimir spectra record the $H$ and $K$ bands.  
These spectra were made with multiple short exposures at 
dithered positions along the slit.  Additional details
are listed in Table~1.  As with the blue and red spectra, 
we obtained flux calibrator spectra at close to the same
time and air mass of the target spectra.  The SpeX 
results were reduced with the Spextool package \citep{cus04}
and the Mimir spectra were extracted using software developed by 
D.\ Clemens\footnote{http://people.bu.edu/clemens/mimir/software.html}.

All the stars observed are targets of continuing programs 
of interferometry with the CHARA Array.  
The targets are listed in Table~2, which gives 
(1) the Henry Draper catalog number, (2) common name, 
(3) spectral classification, (4) stellar effective temperature $T_{\rm eff}$, 
(5) logarithm of the stellar gravity $\log g$, (6) interstellar reddening $E(B-V)$, 
and (7) the HD number of the flux calibrator star adopted.  
The classifications for the Be stars are from the compilation of \citet{yud01}.  
In addition to the Be stars, the list includes three  
Orion supergiants (classifications from \citealt{wal76}),
the luminous blue variable star P~Cygni (classification from 
\citealt{lam83}), the interacting binary $\upsilon$~Sgr 
(classification from \citealt{yud01}), and the yellow supergiant 
$\rho$~Cas (classification from \citealt{bid57}).
The stellar parameters $T_{\rm eff}$ and $\log g$ for the Be stars are taken 
from the apparent values (the average over the visible hemisphere) derived 
by \citet{fre05}, and those for other stars are from the work of 
\citet{sea08} (HD~37128, HD~38771), 
\citet{bou08} (HD~37742),
\citet{dud93} and \citet{leu01} (HD~181615), 
\citet{naj97} (HD~193237), 
\citet{nei05} (HD~202904), and 
\citet{gor06} (HD~224014). 
The reddening estimates $E(B-V)$ are from \citet{dou94} for the Be stars, 
from \citet{shu85} for the O-type stars, and from \citet{dud93}, \citet{naj97}, and 
\citet{zso91} for HD~181615, HD~193237, and HD~224014, respectively.  

\placetable{tab2}      

All the near-IR spectra were transformed to an absolute flux scale 
(and excised of atmospheric telluric features) using the 
{\it xtellcor} software package described by \citet{vac03}.
The method uses flux calibrator stars of spectral classification
A0~V that are transformed to flux through reference to 
a model Vega spectrum calculated by R.\ Kurucz.  
In brief, the procedure involves convolving the model 
Vega spectrum with a kernel designed to match the 
net instrumental and rotational broadening of the 
calibrator spectrum, shifting the model to match 
the Doppler shifted calibrator spectrum, scaling and 
reddening the Vega model to match the calibrator's observed 
magnitudes, dividing the observed calibrator by the model spectrum 
to arrive at a system response spectrum (that includes 
telluric features), and finally dividing the target spectrum
by the response spectrum to obtain a calibrated flux spectrum.  
The transformation is parameterized in the software by setting
the Johnson $B$ and $V$ magnitudes for the calibrator star
(essentially setting the flux zero-point and allowing for
a slope adjustment due to interstellar extinction and/or 
small temperature differences between the calibrator and Vega).   
These $B$ and $V$ magnitudes were selected by making a non-linear, 
least-squares fit of the calibrator $BVJHK_S$ magnitudes to a version of 
the Vega model that was renormalized and reddened for 
interstellar extinction in the same way as done by the 
{\it xtellcor} software.  A small revision of the final $B$ and 
$V$ magnitudes was made to bring the results into consistency
with the recent absolute calibration of IR fluxes by \citet{rie08}. 

The blue and red optical spectra were transformed to
absolute flux following the same basic approach given 
by \citet{vac03}, but the broadening kernel for the 
calibrator was calculated rather than fit using the known 
instrumental broadening and published values of the calibrators' 
projected rotational velocities.  In addition, we also applied a 
small flux correction dependent on the difference in air mass 
between the target and calibrator observations that 
was calculated based upon the mean atmospheric extinction 
coefficients in $BVR$ for KPNO derived by \cite{lan07}.


\section{Optical to Near-IR Spectral Energy Distributions} 

The spectral energy distributions (SEDs) derived for our 24 targets 
are illustrated in Figures 1 -- 3 (in order of increasing HD number).
Each panel shows the flux in a $(\log \lambda, \log \lambda F_\lambda)$ 
format, where $\lambda$ is the wavelength ($\mu$m) and 
$F_\lambda$ is the physical flux received at Earth (in units of 
W~m$^{-2}$~$\mu$m$^{-1}$ = 0.1 erg~cm$^{-2}$~s$^{-1}$~\AA $^{-1}$).  
The first set of measurements from 2006 are 
depicted as black dotted lines, while those for 
2008 are shown over plotted as gray dotted lines.  We calculated 
the average flux over a range of $\pm 0.001 \lambda$ for five 
wavelengths that correspond to line-free regions near the centers 
of the Johnson $BRHKL$ filters, and these mean fluxes appear in 
Table~3.  The errors quoted in Table~3 are the quadratic sum of
several components: (1) instrumental error (primarily from 
Poisson noise detection for these well-exposed spectra), 
(2) repeatability errors (due to fast atmospheric changes between 
the target and calibrator exposures), and (3) errors in 
setting the flux calibration of the calibrator (based upon 
the scatter in the $BVJHK_S$ fit of the calibrator magnitudes). 
In addition, we included a term for the blue and red spectra 
equal to the amount of the extinction correction applied 
to account for air mass mismatch.  We had many cases of multiple 
observations of targets and calibrators that 
we used to estimate the repeatability error ($\approx 4\%$ for the 
blue and red spectra; wavelength dependent for the near-IR spectra, 
but generally just a few percent in wavelength regions with 
good atmospheric transmission).  

\begin{figure}
\begin{center}
\plotone{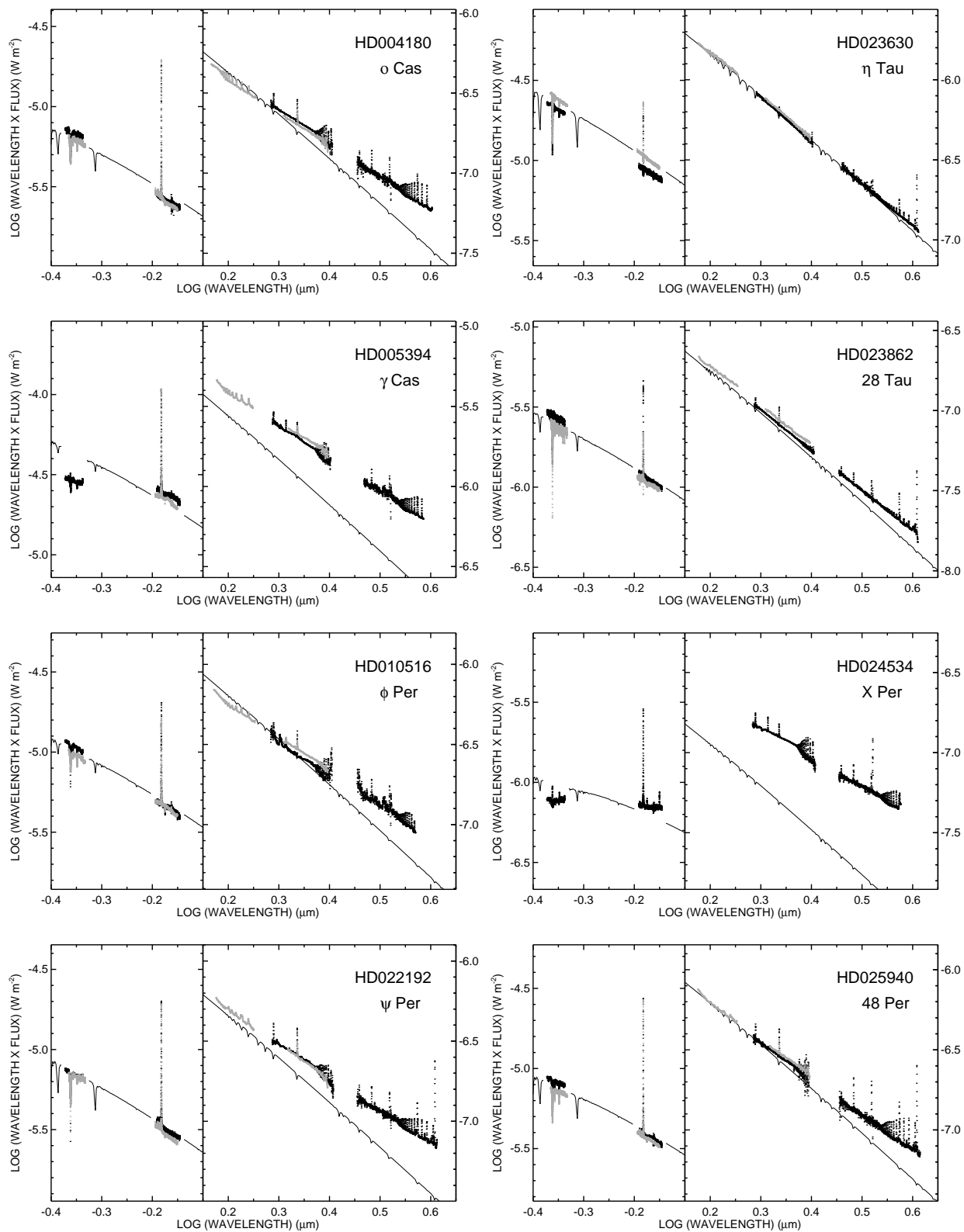}
\end{center}
\caption{The spectral energy distributions of the targets
observed in 2006 ({\it black}) and 2008 ({\it gray}).
The solid lines show the predicted stellar SEDs for 
the parameters from Table~1.}
\label{fig1}
\end{figure}

\placefigure{fig1}     

\begin{figure}
\begin{center}
\plotone{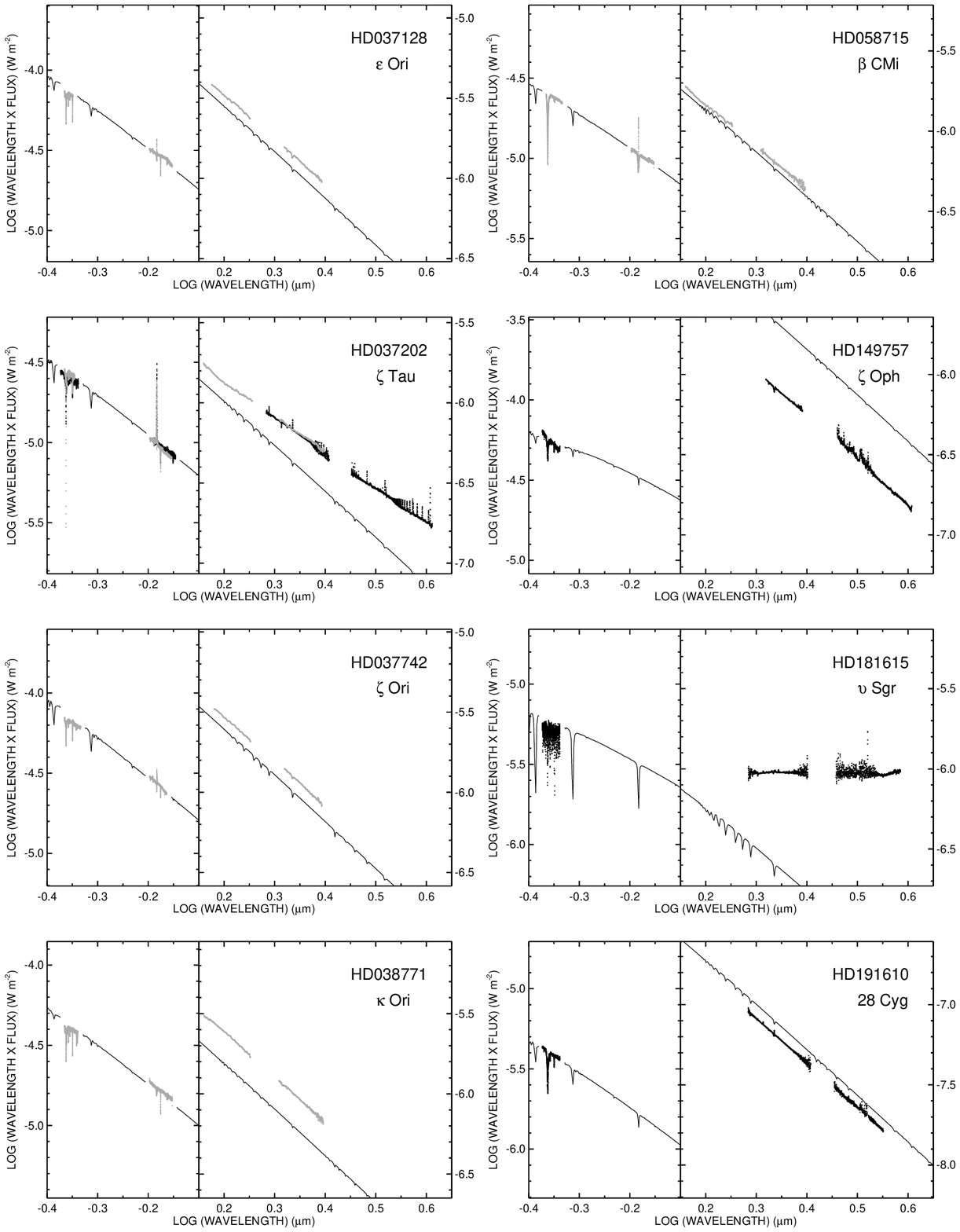}
\end{center}
\caption{The spectral energy distributions in the same format as Fig.~1.}
\label{fig2}
\end{figure}

\placefigure{fig2}     

\begin{figure}
\begin{center}
\plotone{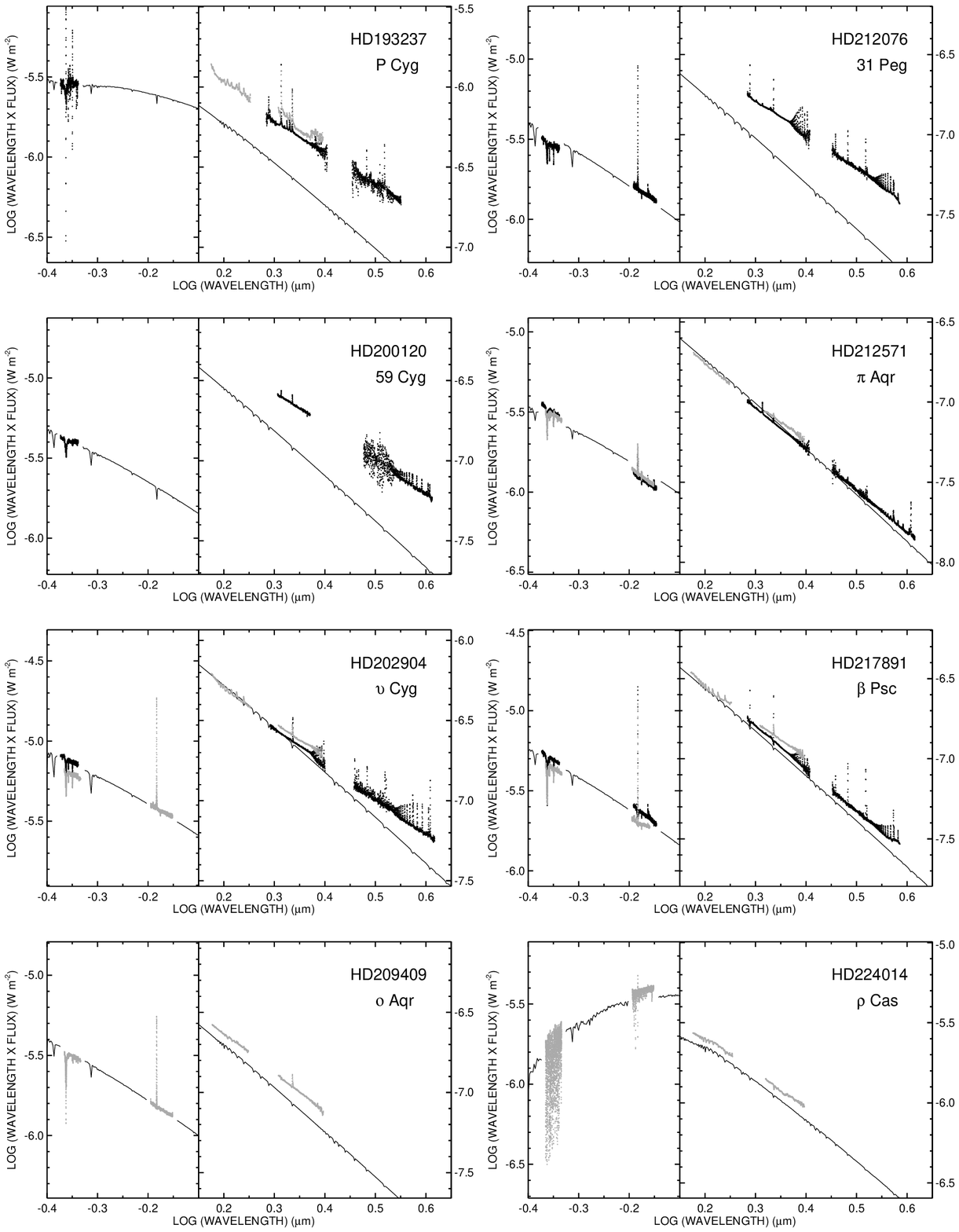}
\end{center}
\caption{The spectral energy distributions in the same format as Fig.~1.}
\label{fig3}
\end{figure}

\placefigure{fig3}     

\placetable{tab3}      

We find that there is overall good agreement between the 
flux levels of the targets between 2006 and 2008, and 
in most cases there is excellent agreement between 
the $K$-band fluxes from the SpeX and Mimir instruments. 
There are a few cases where variability may be 
present ($\eta$~Tau, 28~Tau, 48~Per, P~Cyg, $\upsilon$~Cyg, 
and $\beta$~Psc), and, indeed, the decrease in $B$-band 
flux in 28~Tau = Pleione between 2006 and 2008 occurred
during the development of a new shell phase in that star
\citep{gru07} when additional disk gas projected against the 
star may have caused its optical flux to decline. 
We also note that the Be star HD~24534 = X~Per, the star 
with the largest IR flux excess in our sample (see below), 
was in an historically very bright and strong emission state 
in 2006 \citep{gea07}.  

Any estimate of a flux excess in the SED requires some 
method to determine the stellar contribution across the spectrum. 
We chose to use model spectra based upon the stellar and reddening 
parameters in Table~1 that we normalized to the observed 
fluxes in the blue and red parts of the spectrum.   
We selected solar abundance models with an adopted microturbulence 
parameter of 2 km~s$^{-1}$ that are derived from the grid of line-blanketed, 
local thermodynamical equilibrium (LTE) atmospheres calculated 
by R.\ Kurucz\footnote{http://kurucz.harvard.edu/grids.html}.
These models have limited applicability to our set of targets
for a number of reasons.  The Be stars, for example, are rapid 
rotators with non-spherical shape and with polar regions that are 
hotter than their equatorial zones, and their spectral energy 
distributions probably resemble that of cooler, less luminous
stars \citep{col77}.  Thus, by selecting a temperature and gravity 
that represents some average of the visible hemisphere, the resulting
model spectral energy distribution should be similar (within a few
percent) to the actual flux distribution.  Likewise, the use of LTE versus 
non-LTE models probably introduces a flux error no larger than a few percent 
for the photospheric SEDs of B stars \citep{lh07}.  The greatest 
uncertainty for flux fitting of the Be stars results from the normalization
of the model spectrum in the visible part where the disk contribution 
may be significant (see below).  The supergiants in 
the sample are very luminous and low gravity objects with strong
stellar winds that create an IR-excess \citep{ber85}, and 
a non-LTE and extended atmosphere treatment is necessary.  
Finally, the interacting binary, $\upsilon$~Sgr has a He-enriched atmosphere 
\citep{dud93}.  Despite these various limitations, the LTE flux calculations 
from Kurucz offer an important starting point to search for the deviations in 
the SED that are related to mass outflows and these other effects.

The flux models were calculated by a bilinear interpolation in the 
Kurucz grid using $T_{\rm eff}$ and $\log g$ from Table~1. 
The spectrum was then attenuated for interstellar extinction 
using the reddening from Table~1, a ratio of total-to-selective 
extinction of $3.1$, and the extinction law from 
\citet{fit99}.  The resulting spectrum was normalized by 
(1) smoothing all the available blue and red spectra to the 
spectral resolution of the Kurucz flux spectrum, 
(2) interpolating these to the wavelengths of the model spectrum,
(3) forming an average ratio of observed-to-model flux for 
each spectrum, and (4) determining a global normalization 
factor from the average of all the available spectra. 
This process makes the tacit assumptions that the 
year-to-year flux variations are small (compared to the 
flux calibration errors) and that the stellar component 
dominates the optical flux in the blue and red.  While these
assumptions are reasonable in most cases, we caution that 
the optical flux of the Be stars may be significantly altered
by the presence of a disk.  For example, the optical flux may brighten 
by $\approx 0.5$~mag during emission line outbursts when the disk
is dense \citep{hub98,por03} or may decrease by a comparable 
amount in Be-shell stars with dense disks that block stellar 
flux in the direction of the observer \citep{doa82,hub98}. 
We plan to investigate the extent of the disk contribution 
to the optical spectrum in a subsequent paper that will 
use simple disk density models to fit the interferometric and 
SED observations. 

The model stellar flux distributions are indicated by thin
solid lines in the SED plots of Figures 1 -- 3.  These appear
as continuous in the near-IR range in the right hand panels, 
but they appear only in the gaps between the observed spectral 
regions in the optical region in the left hand panels to 
avoid confusion with the observations.  We find that 
the expected IR flux excess is observed in the SED of most of the 
Be stars (from disks), O-supergiants and P~Cyg (from winds), and 
the interacting binary $\upsilon$~Sgr (from its circumbinary gas).  
We attach no special significance to the relative IR weakness 
observed in the SEDs of HD~149757 and HD~191610 that is probably 
due to somewhat larger errors in their flux calibration.  
We note that the relatively high, red-to-blue flux ratio observed 
in the Be star HD~24534 = X~Per may indicate that a significant 
flux excess is present in the red, so our stellar flux 
normalization made in the optical range may be too high and, 
consequently, the already large IR excess may be underestimated
in this case.   

For the Be stars in our sample with good wavelength coverage, 
we calculated the flux excesses near the centers of the $HKL$ bands 
by determining the ratio of the observed and model average fluxes.
These flux excesses are given in terms of a magnitude difference
in Table~4.  Since these are determined from a normalization 
of the spectrum in the visible region, we symbolically write these as 
$E^\star(Vis-H)$, $E^\star(Vis-K)$, and $E^\star(Vis-L)$, where the 
asterisk is used to differentiate the reddening due to disk emission 
from interstellar reddening (which is accounted for in calculating
the model photospheric spectra).   

\placetable{tab4}      


\section{Emission Line Equivalent-Widths}           

The near-IR spectra of the Be stars contain many emission lines 
formed in the disk, and our data set offers a good opportunity 
to compare the well studied, optical H$\alpha$ emission with 
these near-IR lines in data obtained contemporaneously.  
Figures 4 -- 7 show a montage of (from left to right) the 
H$\alpha$, Br$\gamma$, and Br$\alpha$ $+$ Hu14 emission lines
plus the entire near-IR range, all normalized to the local continuum flux 
(the combined stellar and disk flux).  The $H$-band portions are from 
Mimir data in 2008 and the $KL$-band sections are from the 
2006 SpeX data (high resolution versions).  Tick marks under 
each spectrum show the locations of the H Brackett, Pfund, and Humphreys 
series of lines.  The H$\alpha$ profiles are smoothed to the resolution of
SpeX ($R=2500$).  The H$\alpha$ profiles are from 2006 (2008 for 
$\upsilon$~Cyg), and these include profiles of 28~Cyg and 59~Cyg 
from \citet{gru07} and of P~Cyg from the University of Toledo 
Ritter Observatory (N.\ Richardson et al., in preparation). 
P~Cygni (HD~193237) is not a Be star, but we include it here 
as an example of emission formed in a wind where the gas density 
is lower than typical for Be disks \citep{hon00}.
Our results can be compared directly (in the 2.4 -- 4.1~$\mu$m range)
to spectra from the {\it Infrared Space Observatory} presented 
by \citet{len2a} for HD~5394, 191610, 193237, and 212571.

\begin{figure}
\begin{center}
\plotone{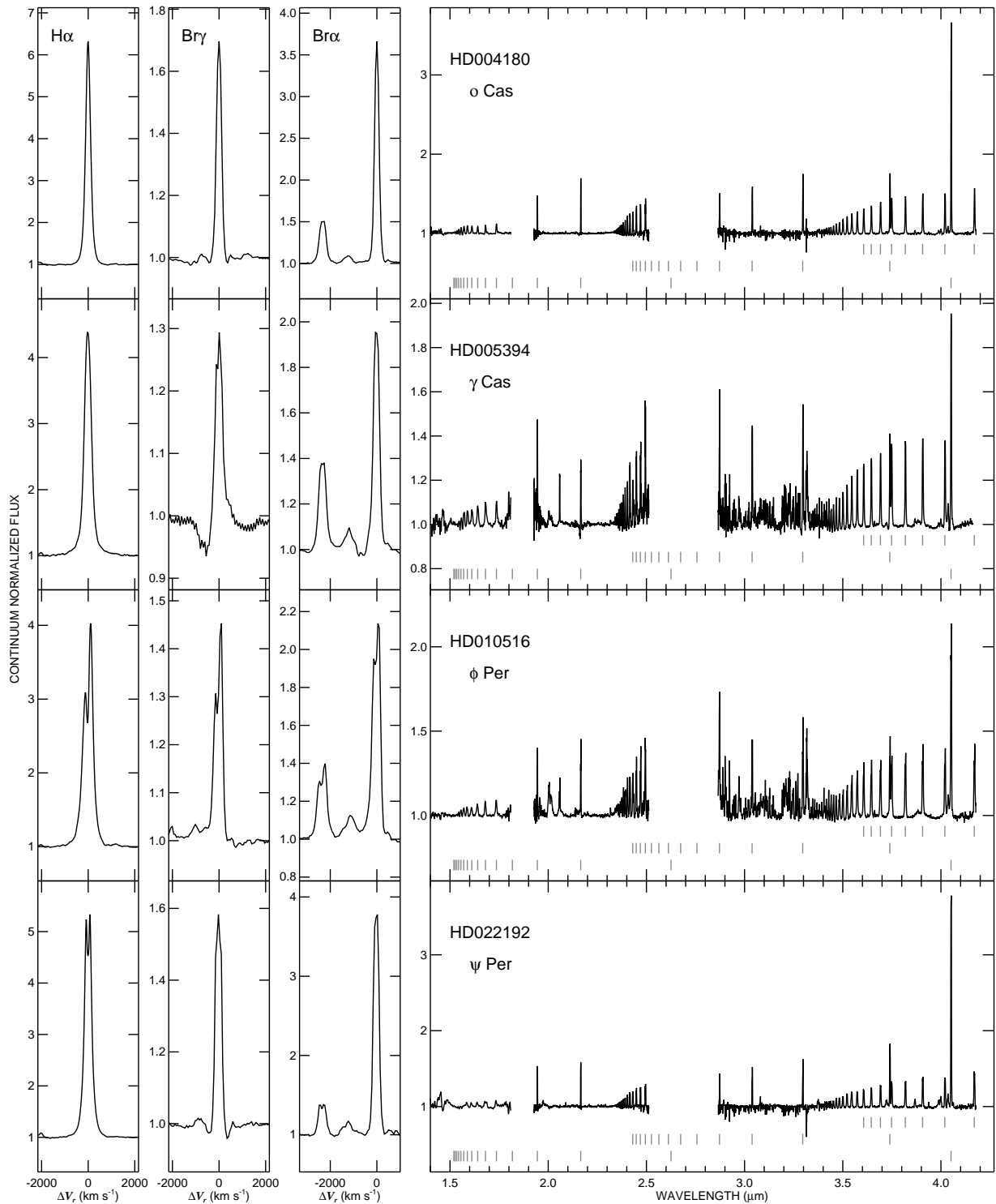}
\end{center}
\caption{The emission line spectra at H$\alpha$, Br$\gamma$, Br$\alpha$,
and across the near-IR region ({\it from left to right}).  
The Hu14 feature appears near $V_r = -2323$ km~s$^{-1}$ in the Br$\alpha$ panel.
Gray tick marks below each spectrum mark the H Brackett, Pfund, 
and Humphreys series ({\it bottom to top}).  The feature at 
2.0 $\mu$m (HD~5394 and 10516) results from incomplete
telluric removal.}
\label{fig4}
\end{figure}

\placefigure{fig4}     

\begin{figure}
\begin{center}
\plotone{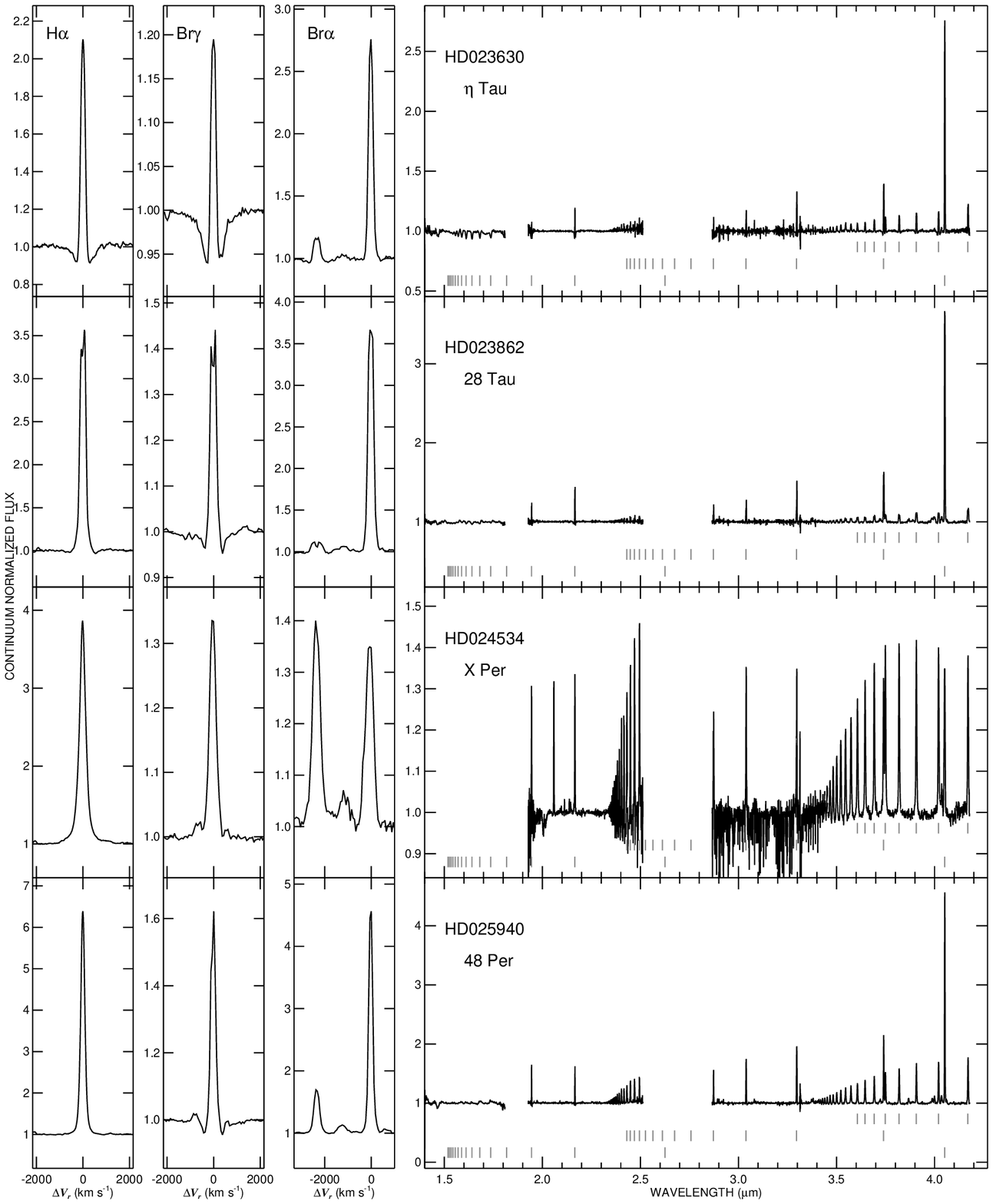}
\end{center}
\caption{The emission line spectra in the same format as Fig.~4.}
\label{fig5}
\end{figure}

\placefigure{fig5}     

\begin{figure}
\begin{center}
\plotone{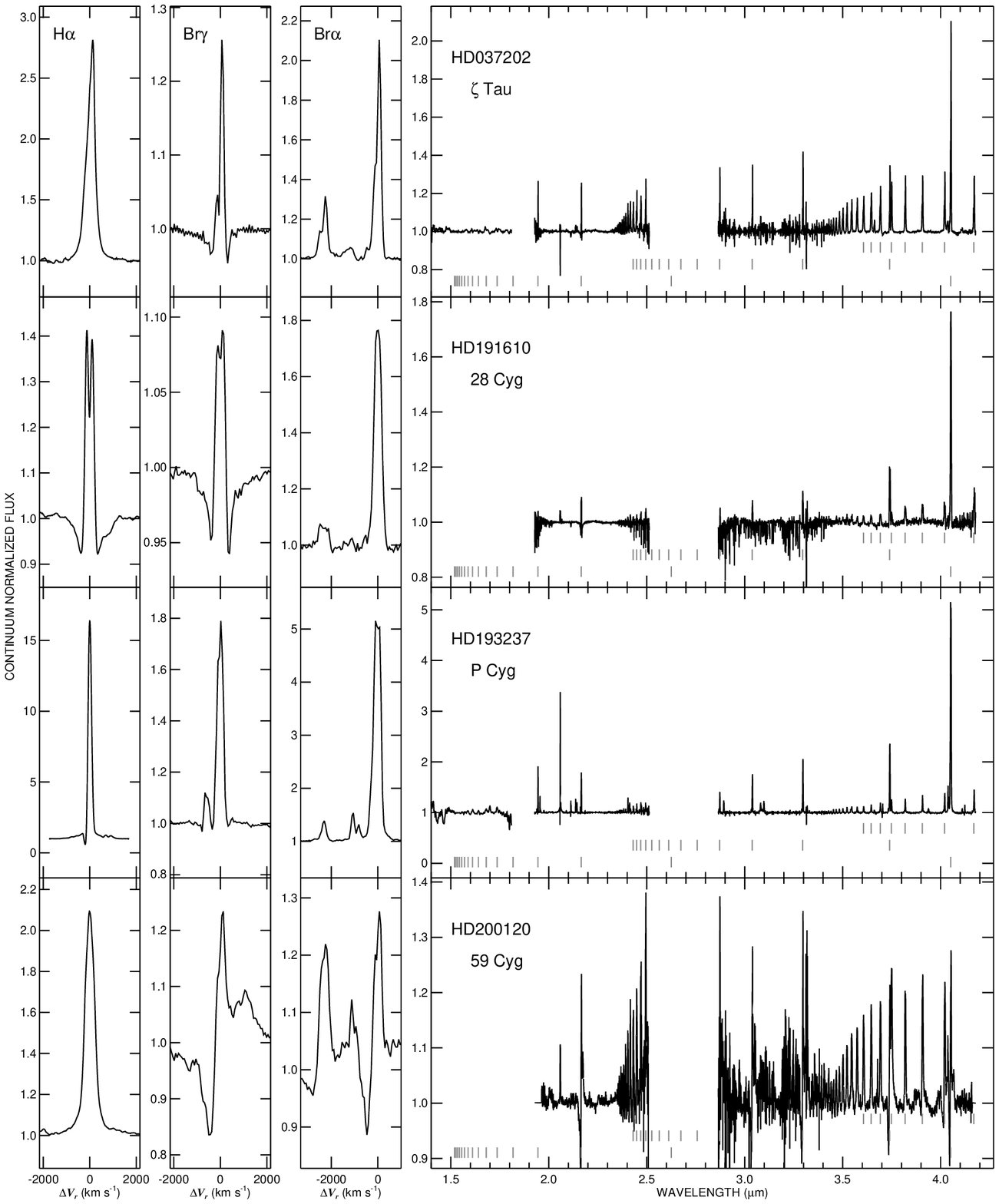}
\end{center}
\caption{The emission line spectra in the same format as Fig.~4.}
\label{fig6}
\end{figure}

\placefigure{fig6}     

\begin{figure}
\begin{center}
\plotone{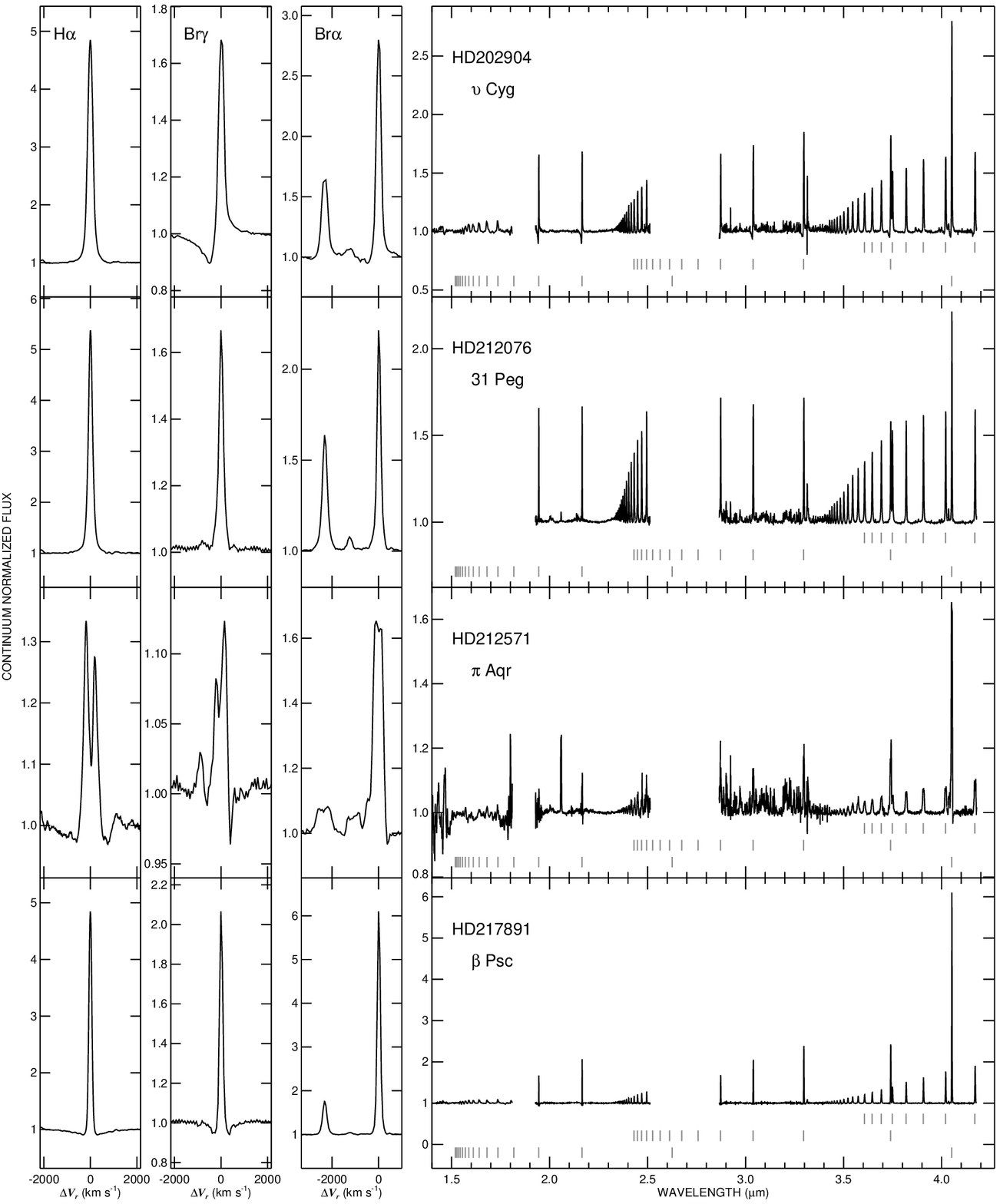}
\end{center}
\caption{The emission line spectra in the same format as Fig.~4.}
\label{fig7}
\end{figure}

\placefigure{fig7}     

We measured the equivalent-width $W_\lambda$ (relative to the local continuum)
by direct integration for the most prominent H emission lines of 
H$\alpha$, Br$\alpha$, Br$\gamma$, Pf$\gamma$, Pf$\delta$, and 
Hu14~$\lambda 4.021$ $\mu$m,  
and these are listed in columns 2 -- 7 of Table~5.  We use the 
standard notation of expressing net emission as a negative 
equivalent-width.  The profiles of Pf$\gamma$ $\lambda 3.741$ $\mu$m 
are partially blended with those of Hu17 $\lambda 3.749$ $\mu$m
in most of the spectra, and we set the upper boundary for 
the integration of Pf$\gamma$ at the minimum position between 
the two features.  The formal measurement errors are approximately
$1\%$ ($3\%$ for Pf$\delta$), but these do not include any 
errors introduced in the telluric removal and flux calibration 
process (where uncertainties in the H line strengths of the 
calibrator spectrum may introduce errors in the final spectrum
of the target).  Furthermore, there were a number of cases 
where the measurement included both emission and absorption 
components.  For example, the Br$\gamma$ profile displayed 
broad photospheric absorption plus narrower disk emission 
for several of the Be stars, and it appeared like a wind feature
with blue absorption plus red emission for P~Cyg, 59~Cyg, and $\upsilon$~Cyg.
In all these cases, the equivalent-width reported in Table~5 
is the net integration of the absorption and emission components. 

\placetable{tab5}      


\section{Discussion}                                

Since both the H emission lines and IR-excess originate in 
the circumstellar disks of Be stars, we might expect that 
the two observables are correlated.  Past work indicates 
that the H$\alpha$ emission strength is related to the 
IR excess.  \citet{kas89} used published data to show that
the H$\alpha$ luminosity is correlated with the IR excess 
luminosity and the spectral type of the underlying star. 
\citet{vke95} obtained near simultaneous H$\alpha$ spectroscopy
and near-IR photometry (to avoid ambiguities introduced 
by time variability of the sources), and they found that 
the equivalent-width of H$\alpha$ was loosely correlated 
with a disk color excess $E^\star(J-L)$, but there was an
intrinsic scatter in the relationship.  These results 
were confirmed (although with somewhat less scatter)
in a study of the relationship of the H$\alpha$ 
equivalent-width and the disk color excess $E^\star(H-K)$ 
by \citet{how01}.  Given the new spectra available from 
our study, we have also explored the relationship between 
the line and continuum emission. 

We need to refer the line emission to the photospheric 
continuum (rather than the observed sum of the photospheric
and disk flux), so we used the flux excess data from Table~4
to derive a line equivalent-width relative to the stellar 
continuum,
\begin{equation}
W_\lambda^\star = W_\lambda ~10^{0.4 E^\star(Vis-\lambda)}
\end{equation}
where $W_\lambda$ is the observed equivalent-width (Table~5)
and $E^\star(Vis-\lambda)$ is the wavelength-interpolated, flux
excess from the disk (Table~4).  We show the derived relationship 
between $W_\lambda^\star$(H$\alpha$) and $E^\star(Vis-L)$ in 
Figure~8.   In this case since the photospheric flux was 
normalized in the optical, there is no net continuum excess 
near H$\alpha$ by definition, and hence $W_\lambda^\star = W_\lambda$.
We see that there is a correlation, but the 
scatter from a one-to-one relationship is significant
(the Spearman's rank correlation coefficient is $\rho = 0.54$). 
We found a similar degree of scatter in plots of 
$W_\lambda^\star$ and $E^\star(Vis-L)$ for 
Br$\alpha$, Br$\gamma$, Pf$\gamma$, and Pf$\delta$, 
and the least scatter is seen in the diagram for 
Hu14~$\lambda 4.021$~$\mu$m (Fig.~9) where the 
Spearman's rank correlation is $\rho = 0.77$. 
The two most discrepant points in Figure~9 (found below
the trend, near $E^\star(Vis-L)=1.0$) correspond to the stars 
$\zeta$~Tau (HD~37202) and 59~Cyg (HD~200120). 
The spectra of both stars show interesting structure 
in the higher resolution SpeX spectra (asymmetric double 
peaks for $\zeta$~Tau and blue absorption for 59~Cyg), 
which suggests that a simple equivalent-width measurement
may be insufficient to explore the relationship between the
line and continuum emission strengths for these two stars.  

\begin{figure}
\begin{center}
{\includegraphics[angle=90,height=12cm]{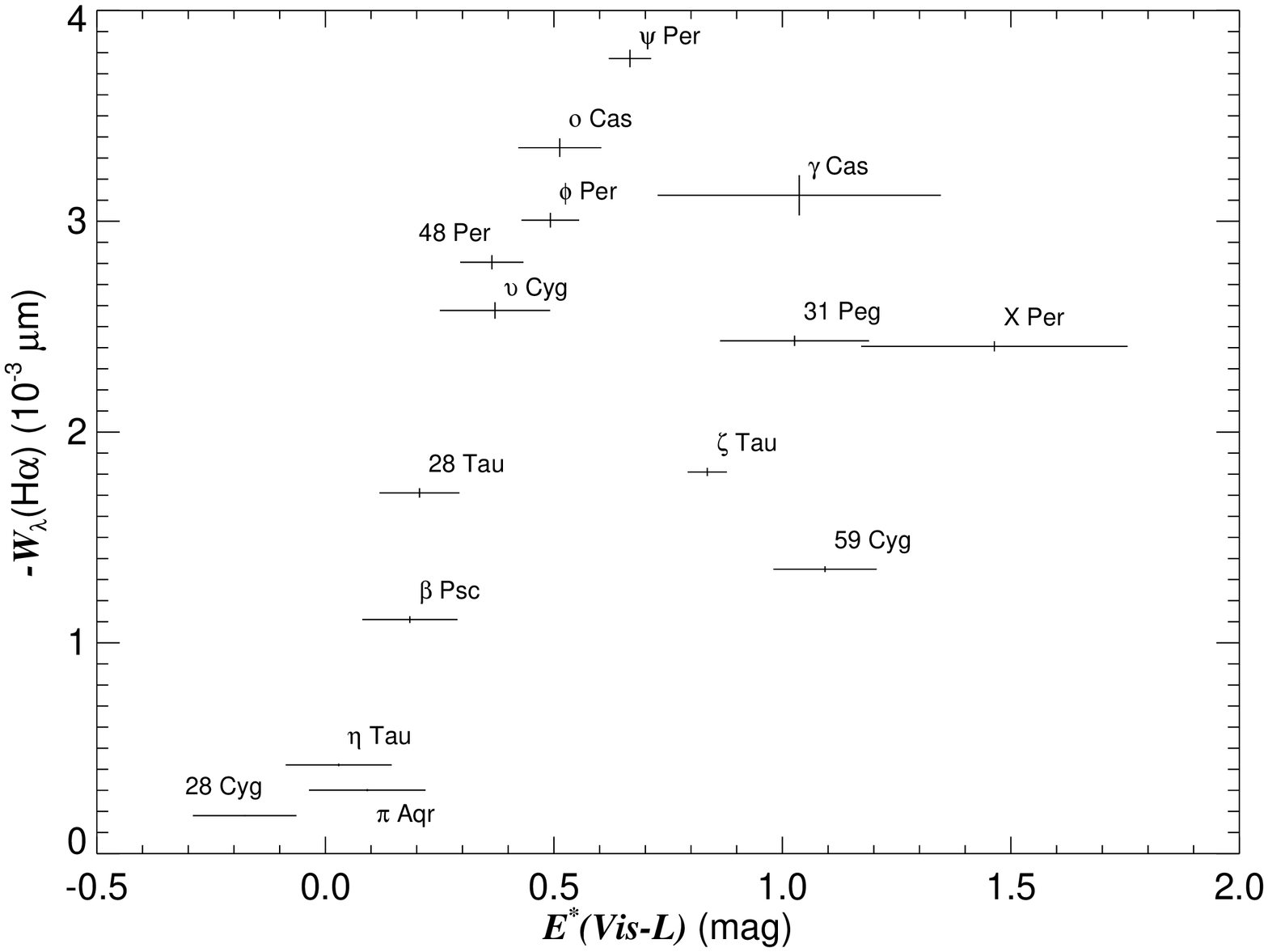}}
\end{center}
\caption{Equivalent-width of the H$\alpha$ emission relative to the 
photospheric continuum plotted against infrared excess $E^\star(Vis-L)$.}
\label{fig8}
\end{figure}

\placefigure{fig8}     

\begin{figure}
\begin{center}
{\includegraphics[angle=90,height=12cm]{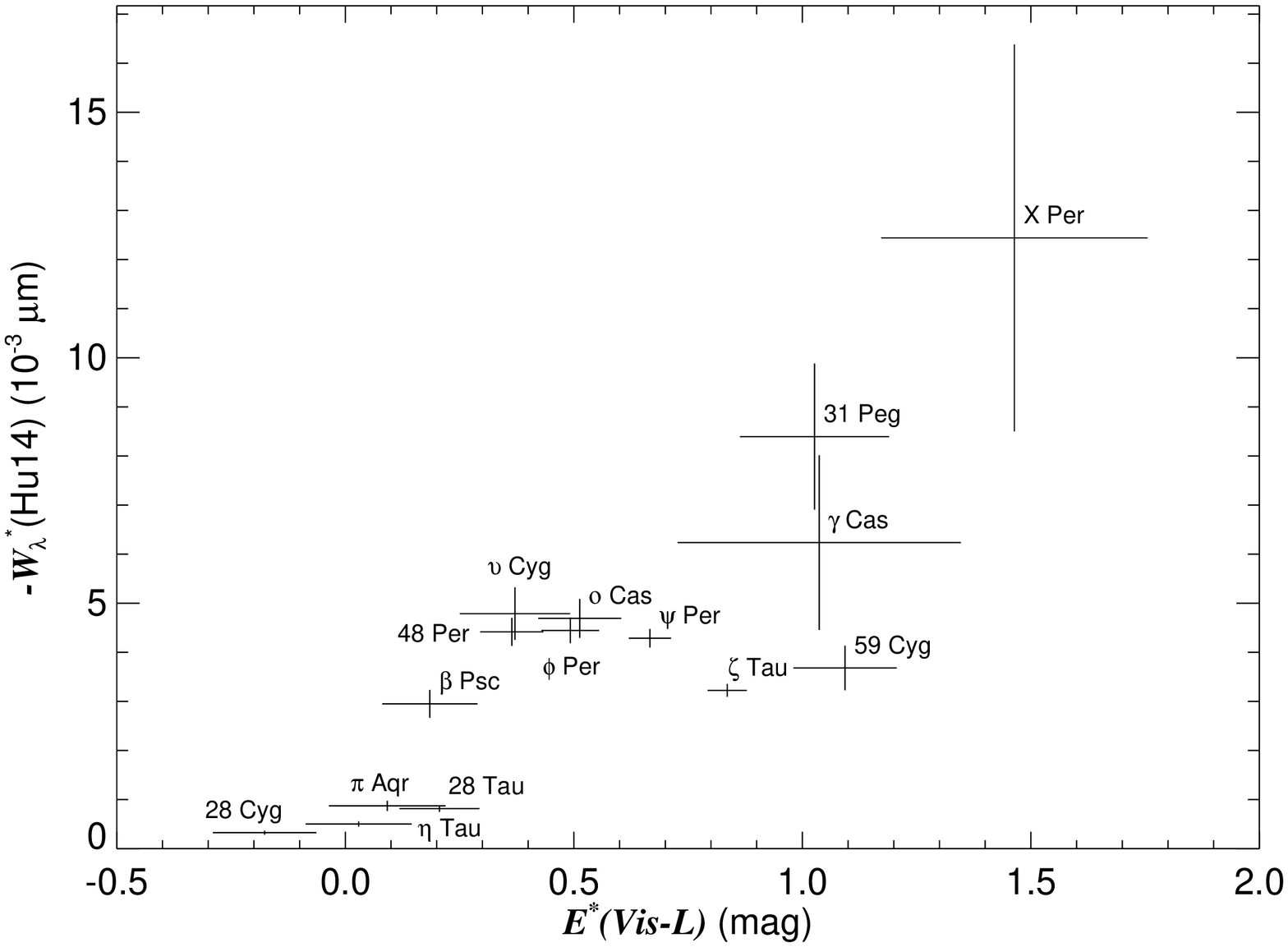}}
\end{center}
\caption{Equivalent-width of the Hu14 emission relative to the 
photospheric continuum plotted against infrared excess $E^\star(Vis-L)$.}
\label{fig9}
\end{figure}

\placefigure{fig9}     

Note that the placement in Figures 8 and 9 may be need to 
be altered if the disk is in fact a significant contributor
to the flux in the visible part of the spectrum (which we 
assumed to be negligible).  Suppose that $\epsilon$ is the ratio
of disk continuum flux to stellar flux in the visible region. 
By assuming $\epsilon=0$, we will underestimate the flux excess
in the IR by an amount 
$\triangle E^\star(Vis-\lambda) \approx 2.5 \log (1 + \epsilon)$
that could be as large as 0.4 mag in cases with large disk 
visible flux contributions.  Furthermore, the H$\alpha$ strength
in Figure~8 would also need to adjusted upwards by a factor 
of $(1 + \epsilon)$ for a measure relative to the stellar continuum. 
These factors may partially explain why the large IR-excess 
cases shown in Figure~8 fall below the expected trend.  

We expect that features of high optical depth will be formed 
over a large range of disk radii (appearing uniformly bright 
over the optically thick regions for an isothermal disk)
while low optical depth features will only appear bright 
in the denser regions of the inner disk.  For example, 
\citet{gie07} found that the angularly resolved disks appear 
smaller in the lower opacity $K$-band continuum compared to
that seen in the high opacity H$\alpha$ line, which is consistent
with the idea that the $K$-band excess forms 
mainly in the inner, denser part of the disk.  Similarly, 
the emission lines of the upper Humphreys series (like Hu14)
are particularly interesting since they probably form mainly in 
the densest region of the disk near the star \citep{hon00,jon09}. 
Since Be disks are time variable and outflowing, we might 
expect that diagnostics that probe like parts of the disk 
will be better correlated than those that form over different ranges of radii.
This expectation agrees with our result that the excess $E^\star(Vis-L)$ 
is better correlated with $W_\lambda^\star$(Hu14) (both forming in the inner disk)
than with $W_\lambda$(H$\alpha$) (which forms out to larger disk radii).

In fact, there is a hint of a better defined relation between 
$W_\lambda$(H$\alpha$) and $E^\star(Vis-L)$ in Figure~8 for $E^\star(Vis-L)<0.7$~mag, 
and the scatter in the relation occurs only for stars with the 
largest IR excesses (densest, largest disks).  Stars with relatively
low density and small disks may have similar radial density 
functions, so that the ratio of flux from H$\alpha$ to that 
in the near-IR continuum is approximately constant.  However, 
the radial density law for Be stars with large, extensive disks 
may be much more complex, reflecting past episodes of differing 
mass loss rates and possibly developing non-axisymmetric structure.
In such a situation, the conditions probed by H$\alpha$ in the 
outer regions may be very different from those in the inner disk 
where the near-IR excess forms.  It is interesting to note that 
all the strong excess stars in Figure~8 (with the exception of 
31~Peg = HD~212076) are known binaries with periods of order 
$\sim 100$~d.  The outer boundaries of the disks in these 
systems are truncated by tidal forces \citep{oka01}, which may 
explain the relative weakness of the H$\alpha$ feature 
compared to the near-IR excess. 

\citet{len2b} suggest that a diagram of the line flux ratios 
of  (log(Hu14/Pf$\gamma$), log(Hu14/Br$\alpha$))
is a useful diagnostic tool to estimate the gas density in 
disks of Be stars \citep{jon09,men09}.  We measured these 
line fluxes by first transforming the continuum of the higher 
resolution SpeX spectra to that of the better flux calibrated 
lower resolution spectra and then subtracting a fit of 
local continuum.  We measured the line fluxes by direct integration 
and then calculated these two line ratios (given in 
columns 8 and 9 of Table~5).  The results are plotted in Figure~10, 
where each symbol is assigned a gray intensity proportional to 
the infrared excess $E^\star(Vis-L)$ (darkest at large $E^\star(Vis-L)$).  
We see a general trend that 
the Be stars in the upper right part of the diagram are those
with the largest IR excess.  As \citet{len2b} point out, 
in optically thick disks, the line flux ratio 
will be given by the product of the line source function ratio
and the ratio of projected radiating surfaces.  
We expect that in very dense environments both these 
ratios will approach unity, so that the ratios for Be stars with 
dense disks will appear in the upper right part of the diagram near 
(log(Hu14/Pf$\gamma$), log(Hu14/Br$\alpha$)) = (0,0), 
the same region where stars with large IR excesses are plotted. 
On the other hand, stars with low density, circumstellar environments 
(like the extended wind of the star P Cygni, 
indicated by a plus sign in Fig.~10) will have line ratios
that tend to populate the lower, left part of the diagram \citep{jon09}. 
However, we suspect that the position in the diagram also 
is modified by the gas temperature in the disk \citep{jon09}, 
since the lowest point in the diagram corresponds to the 
coolest Be star in our sample, 28~Tau (HD~23862; observed 
during a shell phase). 

\begin{figure}
\begin{center}
{\includegraphics[angle=90,height=12cm]{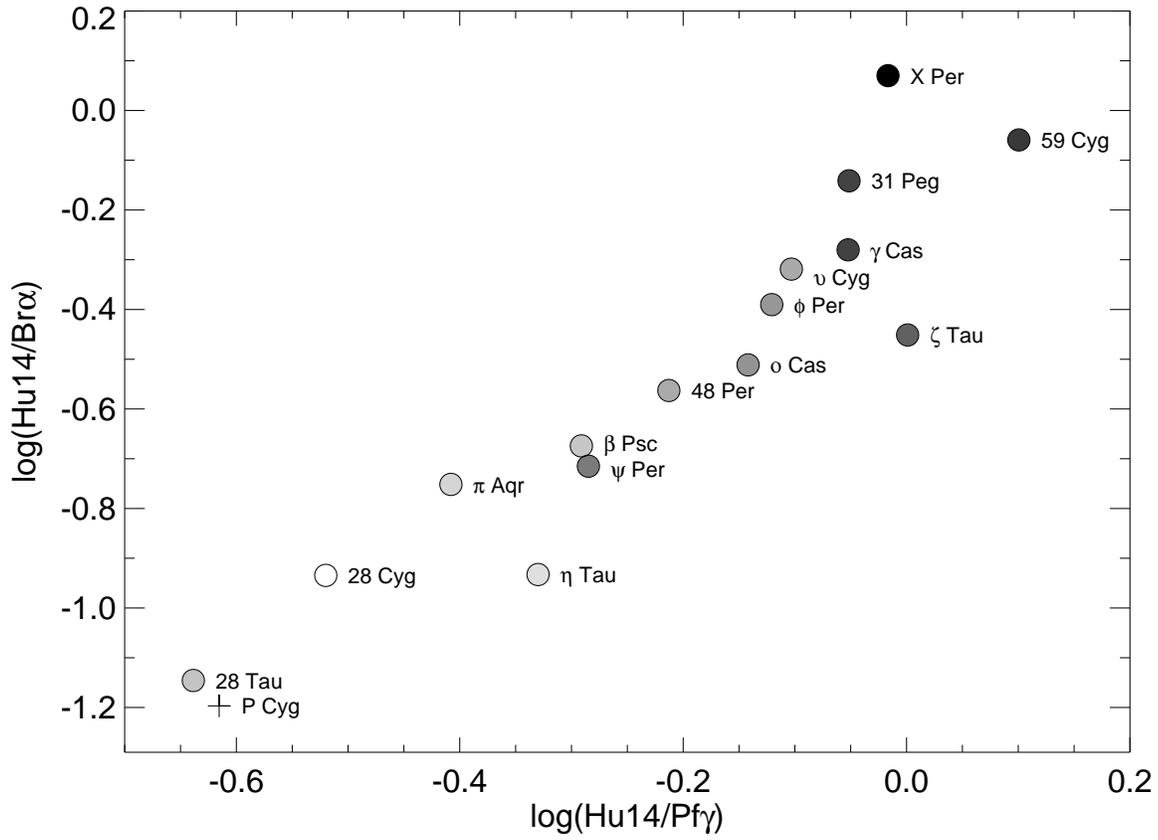}}
\end{center}
\caption{Line flux ratios from Table 5 plotted with a gray 
intensity proportional to the infrared excess $E^\star(Vis-L)$. 
The plus sign indicates the ratios found in the LBV star P~Cyg.}
\label{fig10}
\end{figure}

\placefigure{fig10}    

Curiously, we might expect that in the densest disk regions 
near the star that both the continuum and the Hu14 source 
functions would be similar because they form in regions of 
similar temperature.  If so, then in the optically thick 
parts, the lines would disappear since both lines and continuum
would radiate with the same source function.  Consequently, unless the 
optically thick region of line emission is significantly larger 
than the continuum emitting region, we would expect high excitation
transition lines to vanish in the observed spectrum.  The fact that 
the Hu lines remain as emission features in Be stars with 
dense disks led \citet{hon00} to argue that the lines must 
form in a region of elevated temperature compared to the site of
continuum formation, perhaps in locations above the disk plane. 
Their suggestion appears to be verified in recent models 
of Be disks that show that the mid-plane region is cooler
than off-plane regions in the inner part of the disk
\citep{sig07,car08}.


\section{Conclusions}                               

Our spectrophotometric observations of nearby Be stars 
show that all the stars with strong H$\alpha$ emission 
also display an IR excess relative to the expected 
photospheric flux distribution.  The size of the IR excess 
is correlated with the H$\alpha$ equivalent-width but the 
relation shows the largest scatter among those stars with 
the densest and largest circumstellar disks.  On the 
other hand, the IR excess shows a better correlation with 
the equivalent-widths (corrected for disk continuum emission) 
of high excitation transitions like Hu14.  Since only 
a trace number of H atoms populate these excited states,
transitions like Hu14 have a low opacity except in the 
densest parts of the disk.  We argue that these results 
can be understood in terms of the spatial range in radius 
over which any emission mechanism is optically thick. 
The good correlation between the IR continuum emission 
and the high excitation line emission suggests that both 
form in the inner, dense part of the disk, while the 
less marked correlation between the IR continuum and 
H$\alpha$ emission results from changes in the density 
distribution in the outer part of the disk (perhaps due
to the temporal evolution of the disk and/or the 
tidal influence of a binary companion). 

We are currently making near-IR interferometric observations 
of some 20 northern Be stars with the CHARA Array.  We will
combine the SED data presented here with the interferometric 
visibility data to develop consistent models of the disk 
density structure and orientation in the sky (see our first 
examples in \citealt{gie07}).  


\acknowledgments

We thank Daryl Willmarth and the staff of KPNO and 
John Rayner, Bill Golisch, Dave Griep, and the staff of 
the NASA IRTF for their support of our observational program. 
We also thank Michael Pavel and April Pinnick of 
Boston University for their help with the Mimir data reduction
and an anonymous referee whose comments helped improve the paper. 
Mimir was jointly developed at Boston University and Lowell Observatory 
with support from NASA (NAG5-8716, 9758), NSF (AST-9987335 and AST-0607500), 
and the W.\ M.\ Keck Foundation.  This work was supported by the 
National Science Foundation under grant AST-0606861.
Institutional support has been provided from the GSU College
of Arts and Sciences and from the Research Program Enhancement
fund of the Board of Regents of the University System of Georgia,
administered through the GSU Office of the Vice President
for Research.  




\clearpage

\begin{deluxetable}{lccccl}
\tablewidth{0pc}
\tabletypesize{\scriptsize}
\rotate
\tablenum{1}
\tablecaption{Journal of Spectroscopy \label{tab1}}
\tablehead{
\colhead{UT} &
\colhead{Julian Date} &
\colhead{Wavelength Range} &
\colhead{Resolving Power} &
\colhead{Number of} &
\colhead{Observatory/Telescope/} \\
\colhead{Date} &
\colhead{(HJD-2,450,000)} &
\colhead{($\mu$m)} &
\colhead{($\lambda/\triangle\lambda$)} &
\colhead{Spectra} &
\colhead{Spec., Grating/Detector} \\
\colhead{(1)} &
\colhead{(2)} &
\colhead{(3)} &
\colhead{(4)} &
\colhead{(5)} &
\colhead{(6)}}
\startdata
2006 Sep 15 -- 16&3993.8 -- 3995.1 & 1.92 -- 4.21 &   \phn1700 & 22 & IRTF/3.0m/SpeX, Long XD1.9/Aladdin 3 InSb \\ 
2006 Oct 12      &4020.8 -- 4020.9 & 0.64 -- 0.71 &   \phn6500 & 12 & KPNO/0.9m/Coud\'{e}, B (order 2)/T2KB \\ 
2006 Oct 20 -- 21&4028.6 -- 4029.8 & 0.42 -- 0.46 &   \phn8500 & 18 & KPNO/0.9m/Coud\'{e}, A (order 2)/T2KB \\ 
2008 Oct 18 -- 20&4757.6 -- 4760.0 & 1.40 -- 2.50 &\phn\phn930 & 18 & Lowell/1.8m/Mimir, $JHK$ grism/Aladdin 3 InSb\\
2008 Oct 27 -- 29&4766.6 -- 4769.0 & 0.64 -- 0.71 &      10300 & 17 & KPNO/0.9m/Coud\'{e}, B (order 2)/F3KB \\ 
2008 Nov 30      &4800.6 -- 4801.0 & 0.43 -- 0.46 &      13900 & 16 & KPNO/0.9m/Coud\'{e}, A (order 2)/F3KB \\ 
\enddata
\end{deluxetable}

\clearpage


\begin{deluxetable}{ccccccc}
\tabletypesize{\scriptsize}
\tablewidth{0pt}
\tablenum{2}
\tablecaption{Target and Flux Calibrator Stars\label{tab2}}
\tablehead{
\colhead{Target} &
\colhead{} &
\colhead{Spectral} &
\colhead{$T_{\rm eff}$} &
\colhead{$\log g$} &
\colhead{$E(B-V)$} &
\colhead{Calibrator}  \\
\colhead{HD No.} &
\colhead{Name} &
\colhead{Classification} &
\colhead{(kK)} &
\colhead{(cm s$^{-2}$)} &
\colhead{(mag)} &
\colhead{HD No.}  \\
\colhead{(1)} &
\colhead{(2)} &
\colhead{(3)} &
\colhead{(4)} &
\colhead{(5)} &
\colhead{(6)} &
\colhead{(7)} 
}
\startdata
HD 004180 &  $o$ Cas          & B2 Ve          & 14.4 & 3.3 & 0.11 & HD 001561 \\
HD 005394 &  $\gamma$ Cas     & B0 IVe + sh    & 26.4 & 3.8 & 0.22 & HD 011946 \\
HD 010516 &  $\phi$ Per       & B0.5 IVe + sh  & 25.6 & 3.9 & 0.21 & HD 011946 \\
HD 022192 &  $\psi$ Per       & B4.5 Ve + sh   & 15.8 & 3.5 & 0.11 & HD 025152 \\
HD 023630 &  $\eta$ Tau       & B7 IIIe        & 12.3 & 3.0 & 0.06 & HD 023258 \\
\tablevspace{-5pt}
HD 023862 &  28 Tau           & B8 Vpe + sh    & 12.1 & 3.9 & 0.09 & HD 023258 \\
HD 024534 &  X Per            & O9.5 Vep       & 25.2 & 3.6 & 0.40 & HD 019600 \\
HD 025940 &  48 Per           & B4 Ve          & 16.2 & 3.6 & 0.19 & HD 029526 \\
HD 037128 &  $\epsilon$ Ori   & B0 Ia          & 27.5 & 3.1 & 0.05 & HD 034203 \\
HD 037202 &  $\zeta$ Tau      & B1 IVe + sh    & 19.3 & 3.7 & 0.00 & HD 034203 \\
\tablevspace{-5pt}
HD 037742 &  $\zeta$ Ori      & O9.7 Ib        & 29.5 & 3.3 & 0.04 & HD 034203 \\
HD 038771 &  $\kappa$ Ori     & B0.5 Ia        & 26.0 & 3.0 & 0.04 & HD 045380 \\
HD 058715 &  $\beta$ CMi      & B8 Ve          & 11.8 & 3.8 & 0.01 & HD 060357 \\
HD 149757 &  $\zeta$ Oph      & O9 Ve          & 26.4 & 3.8 & 0.33 & HD 143459 \\
HD 181615 &  $\upsilon$ Sgr   & B2 Vpe         & 11.8 & 2.0 & 0.20 & HD 182678 \\
\tablevspace{-5pt}
HD 191610 &  28 Cyg           & B3 IVe         & 18.4 & 3.7 & 0.06 & HD 192538 \\
HD 193237 &  P Cyg            & B1 Ia+         & 18.2 & 1.2 & 0.51 & HD 192538 \\
HD 200120 &  59 Cyg           & B1.5 Ve + sh   & 21.8 & 3.8 & 0.21 & HD 205314 \\
HD 202904 &  $\upsilon$ Cyg   & B2.5 Vne       & 19.1 & 3.9 & 0.19 & HD 206774 \\
HD 209409 &  $o$ Aqr          & B7 IVe + sh    & 12.9 & 3.7 & 0.05 & HD 212061 \\
\tablevspace{-5pt}
HD 212076 &  31 Peg           & B1.5 Vne       & 19.3 & 3.7 & 0.10 & HD 212061 \\
HD 212571 &  $\pi$ Aqr        & B1 Ve + sh     & 26.1 & 3.9 & 0.22 & HD 212061 \\
HD 217891 &  $\beta$ Psc      & B5 Ve          & 14.4 & 3.7 & 0.05 & HD 212061 \\
HD 224014 &  $\rho$ Cas       & F8 Ia var    &\phn6.0 & 0.7 & 0.42 & HD 223386 \\
\enddata
\end{deluxetable}

\clearpage


\begin{deluxetable}{ccccccc}
\tabletypesize{\scriptsize}
\tablewidth{0pt}
\tablenum{3}
\tablecaption{Monochromatic Fluxes\tablenotemark{a}\label{tab3}}
\tablehead{
\colhead{Star} &
\colhead{Year} &
\colhead{$\log F_\lambda (0.440)$} &
\colhead{$\log F_\lambda (0.680)$} &
\colhead{$\log F_\lambda (1.654)$} &
\colhead{$\log F_\lambda (2.179)$} &
\colhead{$\log F_\lambda (3.410)$}  \\
\colhead{(1)} &
\colhead{(2)} &
\colhead{(3)} &
\colhead{(4)} &
\colhead{(5)} &
\colhead{(6)} &
\colhead{(7)}
}
\startdata
HD004180 &  2006  & \phn$-$8.808 (26)    & \phn$-$9.435 (25)    &       \nodata        &    $-$11.019 (11)    &    $-$11.611 (12)    \\
HD004180 &  2008  & \phn$-$8.843 (27)    & \phn$-$9.445 (24)    &    $-$10.681 (13)    &    $-$11.039 (13)    &       \nodata        \\
\tablevspace{-5pt}
HD005394 &  2006  & \phn$-$8.168 (26)    & \phn$-$8.460 (25)    &       \nodata        &    $-$10.036 (11)    &    $-$10.616 (12)    \\
HD005394 &  2008  &       \nodata        & \phn$-$8.490 (24)    & \phn$-$9.681 (13)    &    $-$10.015 (13)    &       \nodata        \\
\tablevspace{-5pt}
HD010516 &  2006  & \phn$-$8.595 (29)    & \phn$-$9.179 (25)    &       \nodata        &    $-$10.917 (11)    &    $-$11.468 (12)    \\
HD010516 &  2008  & \phn$-$8.646 (28)    & \phn$-$9.187 (24)    &    $-$10.512 (13)    &    $-$10.870 (13)    &       \nodata        \\
\tablevspace{-5pt}
HD022192 &  2006  & \phn$-$8.794 (25)    & \phn$-$9.345 (23)    &       \nodata        &    $-$10.935 (9)\phn &    $-$11.524 (11)    \\
HD022192 &  2008  & \phn$-$8.799 (25)    & \phn$-$9.380 (23)    &    $-$10.569 (12)    &    $-$10.942 (12)    &       \nodata        \\
\tablevspace{-5pt}
HD023630 &  2006  & \phn$-$8.313 (24)    & \phn$-$8.913 (23)    &       \nodata        &    $-$10.553 (9)\phn &    $-$11.271 (10)    \\
HD023630 &  2008  & \phn$-$8.249 (25)    & \phn$-$8.841 (22)    &    $-$10.089 (11)    &    $-$10.530 (11)    &       \nodata        \\
\tablevspace{-5pt}
HD023862 &  2006  & \phn$-$9.207 (24)    & \phn$-$9.794 (23)    &       \nodata        &    $-$11.434 (9)\phn &    $-$12.113 (10)    \\
HD023862 &  2008  & \phn$-$9.261 (24)    & \phn$-$9.819 (22)    &    $-$10.986 (11)    &    $-$11.405 (12)    &       \nodata        \\
\tablevspace{-5pt}
HD024534 &  2006  & \phn$-$9.756 (22)    & \phn$-$9.992 (21)    &       \nodata        &    $-$11.246 (7)\phn &    $-$11.803 (9)\phn \\
\tablevspace{-5pt}
HD025940 &  2006  & \phn$-$8.720 (19)    & \phn$-$9.267 (21)    &       \nodata        &    $-$10.884 (5)\phn &    $-$11.501 (7)\phn \\
HD025940 &  2008  & \phn$-$8.778 (19)    & \phn$-$9.281 (19)    &    $-$10.474 (8)\phn &    $-$10.865 (8)\phn &       \nodata        \\
\tablevspace{-5pt}
HD037128 &  2008  & \phn$-$7.788 (27)    & \phn$-$8.379 (22)    & \phn$-$9.742 (11)    &    $-$10.197 (11)    &       \nodata        \\
\tablevspace{-5pt}
HD037202 &  2006  & \phn$-$8.235 (23)    & \phn$-$8.877 (23)    &       \nodata        &    $-$10.513 (9)\phn &    $-$11.133 (10)    \\
HD037202 &  2008  & \phn$-$8.202 (24)    & \phn$-$8.896 (22)    &    $-$10.135 (11)    &    $-$10.510 (11)    &       \nodata        \\
\tablevspace{-5pt}
HD037742 &  2008  & \phn$-$7.813 (25)    & \phn$-$8.447 (22)    & \phn$-$9.796 (11)    &    $-$10.247 (11)    &       \nodata        \\
\tablevspace{-5pt}
HD038771 &  2008  & \phn$-$8.032 (30)    & \phn$-$8.631 (19)    & \phn$-$9.882 (8)\phn &    $-$10.340 (8)\phn &       \nodata        \\
\tablevspace{-5pt}
HD058715 &  2008  & \phn$-$8.251 (24)    & \phn$-$8.840 (22)    &    $-$10.096 (11)    &    $-$10.536 (11)    &       \nodata        \\
\tablevspace{-5pt}
HD149757 &  2006  & \phn$-$7.893 (99)    &       \nodata        &       \nodata        &    $-$10.424 (7)\phn &    $-$11.150 (9)\phn \\
\tablevspace{-5pt}
HD181615 &  2006  & \phn$-$8.920 (20)    &       \nodata        &       \nodata        &    $-$10.355 (5)\phn &    $-$10.568 (8)\phn \\
\tablevspace{-5pt}
HD191610 &  2006  & \phn$-$9.046 (26)    &       \nodata        &       \nodata        &    $-$11.531 (11)    &    $-$12.264 (13)    \\
\tablevspace{-5pt}
HD193237 &  2006  & \phn$-$9.184 (26)    &       \nodata        &       \nodata        &    $-$10.624 (11)    &    $-$11.204 (12)    \\
HD193237 &  2008  &       \nodata        &       \nodata        &    $-$10.220 (13)    &    $-$10.567 (13)    &       \nodata        \\
\tablevspace{-5pt}
HD200120 &  2006  & \phn$-$9.034 (26)    &       \nodata        &       \nodata        &    $-$10.990 (10)    &    $-$11.566 (12)    \\
\tablevspace{-5pt}
HD202904 &  2006  & \phn$-$8.758 (21)    &       \nodata        &       \nodata        &    $-$10.974 (6)\phn &    $-$11.582 (9)\phn \\
HD202904 &  2008  & \phn$-$8.838 (23)    & \phn$-$9.280 (21)    &    $-$10.562 (9)\phn &    $-$10.936 (9)\phn &       \nodata        \\
\tablevspace{-5pt}
HD209409 &  2008  & \phn$-$9.135 (24)    & \phn$-$9.680 (23)    &    $-$10.895 (12)    &    $-$11.307 (12)    &       \nodata        \\
\tablevspace{-5pt}
HD212076 &  2006  & \phn$-$9.172 (33)    & \phn$-$9.676 (35)    &       \nodata        &    $-$11.201 (9)\phn &    $-$11.805 (10)    \\
\tablevspace{-5pt}
HD212571 &  2006  & \phn$-$9.127 (25)    & \phn$-$9.765 (26)    &       \nodata        &    $-$11.473 (9)\phn &    $-$12.169 (10)    \\
HD212571 &  2008  & \phn$-$9.145 (24)    & \phn$-$9.751 (22)    &    $-$11.026 (12)    &    $-$11.437 (12)    &       \nodata        \\
\tablevspace{-5pt}
HD217891 &  2006  & \phn$-$8.935 (31)    & \phn$-$9.486 (41)    &       \nodata        &    $-$11.230 (9)\phn &    $-$11.941 (10)    \\
HD217891 &  2008  & \phn$-$8.991 (27)    & \phn$-$9.548 (22)    &    $-$10.819 (12)    &    $-$11.193 (12)    &       \nodata        \\
\tablevspace{-5pt}
HD224014 &  2008  & \phn$-$9.559 (19)    & \phn$-$9.242 (20)    & \phn$-$9.843 (8)\phn &    $-$10.233 (8)\phn &       \nodata        \\
\enddata
\tablenotetext{a}{$\log_{10}$ of the observed flux in units of W m$^{-2}$ $\mu$m$^{-1}$ at 
wavelengths specified in microns.  Numbers in parentheses give the errors in units of the last 
digit quoted.}
\end{deluxetable}

\clearpage


\begin{deluxetable}{ccccc}
\tabletypesize{\small}
\tablewidth{0pt}
\tablenum{4}
\tablecaption{Be Star Flux Excess\tablenotemark{a}\label{tab4}}
\tablehead{
\colhead{} &
\colhead{} &
\colhead{$E^\star(Vis-H)$} &
\colhead{$E^\star(Vis-K)$} &
\colhead{$E^\star(Vis-L)$} \\ 
\colhead{Star} &
\colhead{Year} &
\colhead{(mag)} &
\colhead{(mag)} &
\colhead{(mag)}  \\
\colhead{(1)} &
\colhead{(2)} &
\colhead{(3)} &
\colhead{(4)} &
\colhead{(5)} 
}
\startdata
HD004180 &  2006  &    \nodata       & \phs0.15 (9)\phn & \phs0.51 (9)\phn \\
HD004180 &  2008  &  $-$0.12 (9)\phn & \phs0.10 (9)\phn &    \nodata       \\
\tablevspace{-5pt}
HD005394 &  2006  &    \nodata       & \phs0.61 (31)    & \phs1.04 (31)    \\
HD005394 &  2008  & \phs0.36 (31)    & \phs0.66 (31)    &    \nodata       \\
\tablevspace{-5pt}
HD010516 &  2006  &    \nodata       &  $-$0.01 (6)\phn & \phs0.49 (6)\phn \\
HD010516 &  2008  &  $-$0.13 (6)\phn & \phs0.11 (6)\phn &    \nodata       \\
\tablevspace{-5pt}
HD022192 &  2006  &    \nodata       & \phs0.30 (4)\phn & \phs0.67 (5)\phn \\
HD022192 &  2008  & \phs0.09 (5)\phn & \phs0.28 (5)\phn &    \nodata       \\
\tablevspace{-5pt}
HD023630 &  2006  &    \nodata       & \phs0.00 (12)    & \phs0.03 (12)    \\
HD023630 &  2008  & \phs0.03 (12)    & \phs0.05 (12)    &    \nodata       \\
\tablevspace{-5pt}
HD023862 &  2006  &    \nodata       & \phs0.07 (9)\phn & \phs0.21 (9)\phn \\
HD023862 &  2008  & \phs0.07 (9)\phn & \phs0.14 (9)\phn &    \nodata       \\
\tablevspace{-5pt}
HD024534 &  2006  &    \nodata       & \phs1.01 (29)    & \phs1.46 (29)    \\
\tablevspace{-5pt}
HD025940 &  2006  &    \nodata       & \phs0.07 (7)\phn & \phs0.36 (7)\phn \\
HD025940 &  2008  &  $-$0.01 (7)\phn & \phs0.12 (7)\phn &    \nodata       \\
\tablevspace{-5pt}
HD037202 &  2006  &    \nodata       & \phs0.51 (4)\phn & \phs0.84 (4)\phn \\
HD037202 &  2008  & \phs0.30 (4)\phn & \phs0.52 (4)\phn &    \nodata       \\
\tablevspace{-5pt}
HD058715 &  2008  & \phs0.08 (4)\phn & \phs0.10 (4)\phn &    \nodata       \\
\tablevspace{-5pt}
HD191610 &  2006  &    \nodata       &  $-$0.21 (11)    &  $-$0.18 (11)    \\
\tablevspace{-5pt}
HD200120 &  2006  &    \nodata       & \phs0.68 (11)    & \phs1.09 (11)    \\
\tablevspace{-5pt}
HD202904 &  2006  &    \nodata       & \phs0.05 (12)    & \phs0.37 (12)    \\
HD202904 &  2008  &  $-$0.04 (12)    & \phs0.14 (12)    &    \nodata       \\
\tablevspace{-5pt}
HD209409 &  2008  & \phs0.18 (7)\phn & \phs0.27 (7)\phn &    \nodata       \\
\tablevspace{-5pt}
HD212076 &  2006  &    \nodata       & \phs0.68 (16)    & \phs1.03 (16)    \\
\tablevspace{-5pt}
HD212571 &  2006  &    \nodata       &  $-$0.05 (13)    & \phs0.09 (13)    \\
HD212571 &  2008  &  $-$0.06 (13)    & \phs0.04 (13)    &    \nodata       \\
\tablevspace{-5pt}
HD217891 &  2006  &    \nodata       & \phs0.11 (10)    & \phs0.18 (10)    \\
HD217891 &  2008  & \phs0.01 (10)    & \phs0.20 (11)    &    \nodata       \\
\enddata
\tablenotetext{a}{Numbers in parentheses give the errors in units of the last 
digit quoted.}
\end{deluxetable}


\begin{deluxetable}{ccccccccc}
\tabletypesize{\scriptsize}
\rotate
\tablewidth{0pt}
\tablenum{5}
\tablecaption{Be Star Line Equivalent-Widths and Ratios\label{tab5}}
\tablehead{
\colhead{} &
\colhead{$-W_\lambda$(H$\alpha$)} &
\colhead{$-W_\lambda$(Br$\alpha$)} &
\colhead{$-W_\lambda$(Br$\gamma$)} &
\colhead{$-W_\lambda$(Pf$\gamma$)} &
\colhead{$-W_\lambda$(Pf$\delta$)} &
\colhead{$-W_\lambda$(Hu14)} &
\colhead{} &
\colhead{} \\ 
\colhead{Star} &
\colhead{($10^{-3}$ $\mu$m)} &
\colhead{($10^{-3}$ $\mu$m)} &
\colhead{($10^{-3}$ $\mu$m)} &
\colhead{($10^{-3}$ $\mu$m)} &
\colhead{($10^{-3}$ $\mu$m)} &
\colhead{($10^{-3}$ $\mu$m)} &
\colhead{$\log$(Hu14/Pf$\gamma$)} &
\colhead{$\log$(Hu14/Br$\alpha$)}  \\
\colhead{(1)} &
\colhead{(2)} &
\colhead{(3)} &
\colhead{(4)} &
\colhead{(5)} &
\colhead{(6)} &
\colhead{(7)} &
\colhead{(8)} &
\colhead{(9)}
}
\startdata
HD004180&     3.35&\phn     8.77&    \phs 1.33&     2.96&     2.27&     2.58
& $-$0.14& $-$0.51 \\
HD005394&     3.12&\phn     4.35&    \phs 0.55&     1.94&     2.15&     2.10
& $-$0.05& $-$0.28 \\
HD010516&     3.01&\phn     6.35&    \phs 1.18&     2.43&     2.75&     2.44
& $-$0.12& $-$0.39 \\
HD022192&     3.77&        11.34&    \phs 1.23&     3.45&     1.80&     2.05
& $-$0.28& $-$0.71 \\
HD023630&     0.42&\phn     5.70&      $-$0.02&     1.20&     0.76&     0.49
& $-$0.33& $-$0.93 \\
\tablevspace{-5pt}
HD023862&     1.71&        11.21&    \phs 0.76&     2.68&     1.31&     0.65
& $-$0.64& $-$1.15 \\
HD024534&     2.41&\phn     2.05&    \phs 0.95&     1.87&     1.20&     2.35
& $-$0.02&\phs0.07 \\
HD025940&     2.81&        10.83&    \phs 0.98&     3.73&     3.04&     2.89
& $-$0.21& $-$0.56 \\
HD037202&     1.81&\phn     4.00&    \phs 0.20&     1.08&     0.89&     1.34
&\phs0.00& $-$0.45 \\
HD191610&     0.18&\phn     4.17&      $-$0.03&     1.09&     0.36&     0.39
& $-$0.52& $-$0.93 \\
\tablevspace{-5pt}
HD193237\tablenotemark{a}&6.66&22.85&\phs 1.89&     5.05&     3.47&     1.36
& $-$0.62& $-$1.20 \\
HD200120&     1.35&\phn     1.15&    \phs 0.08&     0.66&     0.75&     1.05
&\phs0.10& $-$0.06 \\
HD202904&     2.58&\phn     6.48&    \phs 1.06&     2.94&     2.73&     3.10
& $-$0.10& $-$0.32 \\
HD212076&     2.43&\phn     3.72&    \phs 1.35&     2.28&     2.56&     2.60
& $-$0.05& $-$0.14 \\
HD212571&     0.30&\phn     4.75&    \phs 0.41&     1.51&     1.13&     0.78
& $-$0.41& $-$0.75 \\
\tablevspace{-5pt}
HD217891&     1.11&        12.12&    \phs 1.07&     3.84&     3.57&     2.46
& $-$0.29& $-$0.67 \\
\enddata
\tablenotetext{a}{P Cygni is a Luminous Blue Variable and not a Be star, but it 
is included here for comparison.}
\end{deluxetable}

\clearpage



\clearpage

\input{epsf}

\clearpage

\clearpage

\clearpage

\clearpage

\clearpage

\clearpage

\clearpage

\clearpage

\clearpage

\clearpage



\clearpage

\begin{thebibliography}{}
\bibitem[Bertout et al.(1985)]{ber85}
         Bertout, C., Leitherer, C., Stahl, O., \& Wolf, B. 1985, \aap, 144, 87
\bibitem[Bidelman \& McKellar(1957)]{bid57}
         Bidelman, W. P., \& McKellar, A. 1957, \pasp, 69, 31
\bibitem[Bouret et al.(2008)]{bou08}
         Bouret, J.-C., Donati, J.-F., Martins, F., Escolano, C.,
         Marcolino, W., Lanz, T., \& Howarth, I. D. 2008, \mnras, 389, 75
\bibitem[Carciofi \& Bjorkman(2008)]{car08}
         Carciofi, A. C., \& Bjorkman, J. E. 2008, \apj, 684, 1374
\bibitem[Carciofi et al.(2009)]{car09}
         Carciofi, A. C., Okazaki, A. T., le Bouquin, J.-B., 
         \v{S}tefl, S., Rivinius, T., Baade, D., Bjorkman, J. E., \&
         Hummel, C. A. 2009, \aap, in press (arXiv:0901.1098)
\bibitem[Clark \& Steele(2000)]{cla00}
         Clark, J. S., \& Steele, I. A. 2000, \aaps, 141, 65
\bibitem[Clemens et al.(2007)]{cle07}
         Clemens, D. P., Sarcia, D., Grabau, A., Tollestrup, E. V., 
         Buie, M. W., Dunham, E., \& Taylor, B.
         2007, \pasp, 119, 1385
\bibitem[Collins \& Sonneborn(1977)]{col77}
         Collins, G. W., II, \& Sonneborn, G. H. 1977, \apjs, 34, 41
\bibitem[Cushing et al.(2004)]{cus04}
         Cushing, M. C., Vacca, W. D., \& Rayner, J. T. 2004, \pasp, 116, 362
\bibitem[Doazan(1982)]{doa82}
         Doazan, V. 1982, in B Stars with and without Emission Lines, 
         ed. A. Underhill \& V. Doazan (Washington, DC: NASA SP-456), 279
\bibitem[Dougherty et al.(1994)]{dou94}
         Dougherty, S. M., Waters, L. B. F. M., Burki, G., Cot\'{e}, J.,
         Cramer, N., van Kerkwijk, M. H., \& Taylor, A. R.
         1994, \aap, 290, 609
\bibitem[Dudley \& Jeffery(1993)]{dud93}
         Dudley, R. E., \& Jeffery, C. S. 1993, \mnras, 262, 945
\bibitem[Fitzpatrick(1999)]{fit99}
         Fitzpatrick, E. L. 1999, \pasp, 111, 63
\bibitem[Fr\'{e}mat et al.(2005)]{fre05}
         Fr\'{e}mat Y., Zorec J., Hubert A.-M., \& Floquet M. 2005, \aap, 440, 305
\bibitem[Gies et al.(2007)]{gie07}
         Gies, D. R., et al. 2007, \apj, 654, 527
\bibitem[Gorlova et al.(2006)]{gor06}
         Gorlova, N., Lobel, A., Burgasser, A. J., Rieke, G. H., Ilyin, I., \&
         Stauffer, J. R. 2006, \apj, 651, 1130
\bibitem[Grundstrom(2007)]{gru07}
         Grundstrom, E. D. 2007, Ph.D. dissertation (Georgia State Univ.)
\bibitem[Grundstrom et al.(2007)]{gea07}
         Grundstrom, E. D., et al. 2007, \apj, 660, 1398
\bibitem[Hony et al.(2000)]{hon00}
         Hony, S., et al. 2000, \aap, 355, 187
\bibitem[Howells et al.(2001)]{how01}
         Howells, L., Steele, I. A., Porter, J. M., \& Etherton, J.
         2001, \aap, 369, 99
\bibitem[Hubert \& Floquet(1998)]{hub98}
         Hubert, A. M., \& Floquet, M. 1998, \aap, 335, 565
\bibitem[Jones et al.(2009)]{jon09}
         Jones, C. E., Molak, A., Sigut, T. A. A., de Koter, A.,
         Lenorzer, A., \& Popa, S. C. 2009, \mnras, 392, 383
\bibitem[Kastner \& Mazzali(1989)]{kas89}
         Kastner, J. H., \& Mazzali, P. A. 1989, \aap, 210, 295
\bibitem[Lamers et al.(1983)]{lam83}
         Lamers, H. J. G. L. M., de Groot, M., \& Cassatella, A.
         1983, \aap, 128, 299
\bibitem[Landolt \& Uomoto(2007)]{lan07}
         Landolt, A. U., \& Uomoto, A. K. 2007, \aj, 133, 768
\bibitem[Lanz \& Hubeny(2007)]{lh07}
         Lanz, T., \& Hubeny, I. 2007, \apjs, 169, 83
\bibitem[Lenorzer et al.(2002b)]{len2b}
         Lenorzer, A., de Koter, A., \& Waters, L. B. F. M.
         2002b, \aap, 386, L5
\bibitem[Lenorzer et al.(2002a)]{len2a}
         Lenorzer, A., Vandenbussche, B., Morris, P., de Koter, A.,
         Geballe, T. R., Waters, L. B. F. M., Hony, S., \& Kaper, L.
         2002a, \aap, 384, 473
\bibitem[Leushin(2001)]{leu01}
         Leushin, V. V. 2001, Astron. Lett., 27, 634
\bibitem[McSwain et al.(2008)]{mcs08}
         McSwain, M. V., Huang, W., Gies, D. R., Grundstrom, E. D., \&
         Townsend, R. H. D. 2008, \apj, 672, 590
\bibitem[Meilland et al.(2007)]{mei07}
         Meilland, A., et al. 2007, \aap, 464, 59
\bibitem[Mennickent et al.(2009)]{men09}
         Mennickent, R. E., Sabogal, B., Granada, A., \& Cidale, L.
         2009, \pasp, 121, 125
\bibitem[Najarro et al.(1997)]{naj97}
         Najarro, F., Hillier, D. J., \& Stahl, O. 1997, \aap, 326, 1117	
\bibitem[Neiner et al.(2005)]{nei05}
         Neiner, C., et al. 2005, \aap, 437, 257
\bibitem[Okazaki \& Negueruela(2001)]{oka01}
         Okazaki, A. T., \& Negueruela, I. 2001, \aap, 377, 161
\bibitem[Porter(1999)]{por99}
         Porter, J. M. 1999, \aap, 348, 512
\bibitem[Porter \& Rivinius(2003)]{por03}
         Porter, J. M., \& Rivinius, Th. 2003, \pasp, 115, 1153
\bibitem[Rayner et al.(2003)]{ray03}
         Rayner, J. T., Toomey, D. W., Onaka, P. M., Denault, A. J.,
         Stahlberger, W. E., Vacca, W. D., Cushing, M. C., \& Wang, S.
         2003, \pasp, 115, 362
\bibitem[Rieke et al.(2008)]{rie08}
         Rieke, G. H., et al. 2008, \aj, 135, 2245
\bibitem[Searle et al.(2008)]{sea08}
         Searle, S. C., Prinja, R. K., Massa, D., \& Ryans, R.
         2008, \aap, 481, 777
\bibitem[Shull \& van Steenberg(1985)]{shu85}
         Shull, J. M., \& van Steenberg, M. E. 
         1985, \apj, 294, 599
\bibitem[Sigut \& Jones(2007)]{sig07}
         Sigut, T. A. A., \& Jones, C. E. 2007, \apj, 668, 481
\bibitem[Stee \& Bittar(2001)]{ste01}
         Stee, P., \& Bittar, J. 2001, \aap, 367, 532
\bibitem[Steele \& Clark(2001)]{stl01}
         Steele, I. A., \& Clark, J. S. 2001, \aap, 371, 643
\bibitem[ten Brummelaar et al.(2005)]{ten05}
         ten Brummelaar, T. A., et al. 2005, \apj, 628, 453
\bibitem[Vacca et al.(2003)]{vac03}
         Vacca, W. D., Cushing, M. C., \& Rayner, J. T. 2003, \pasp, 115, 389
\bibitem[van Kerkwijk et al.(1995)]{vke95}
         van Kerkwijk, M. H., Waters, L. B. F. M., \& Marlborough, J. M.
         1995, \aap, 300, 259
\bibitem[Walborn(1976)]{wal76}
         Walborn, N. R. 1976, \apj, 205, 419
\bibitem[Waters(1986)]{wat86}
         Waters, L. B. F. M. 1986, \aap, 162, 121
\bibitem[Yudin(2001)]{yud01}
         Yudin, R. V. 2001, \aap, 368, 912
\bibitem[Zsoldos \& Percy(1991)]{zso91}
         Zsoldos, E., \& Percy, J. R. 1991, \aap, 246, 441
\end{thebibliography}
\end{document}